\documentclass[journal abbreviation, manuscript]{copernicus}
\usepackage{subfig} %should come after caption package if given, can't be used with caption2, defines subfloat
\usepackage{caption}
\usepackage{xcolor}
\usepackage{url}
\usepackage{float}
\usepackage{algorithm} 
\usepackage{algorithmic}
\usepackage{booktabs}
\usepackage{multirow}
\usepackage{makecell}
\usepackage{stackengine}
\usepackage{wrapfig}
\usepackage{lineno}
\nolinenumbers

\begin{document}

\title{Estimating Solar Irradiance Using Sky Imagers}

\Author[1,2]{Soumyabrata}{Dev}
\Author[3]{Florian M.}{Savoy}
\Author[4]{Yee Hui}{Lee}
\Author[3,5]{Stefan}{Winkler}

\affil[1]{ADAPT SFI Research Centre, Dublin, Ireland}
\affil[2]{School of Computer Science, University College Dublin, Ireland}
\affil[3]{Advanced Digital Sciences Center (ADSC), University of Illinois at Urbana-Champaign, Singapore 138632}
\affil[4]{School of Electrical and Electronic Engineering, Nanyang Technological University (NTU), Singapore 639798}
\affil[5]{School of Computing, National University of Singapore (NUS), Singapore 117417}

\runningtitle{TEXT}

\runningauthor{TEXT}

\correspondence{Stefan Winkler (winkler@comp.nus.edu.sg)}

\received{}
\pubdiscuss{} %% only important for two-stage journals
\revised{}
\accepted{}
\published{}

%% These dates will be inserted by Copernicus Publications during the typesetting process.

\firstpage{1}

\maketitle

\begin{abstract}
Ground-based whole sky cameras are extensively used for localized monitoring of clouds nowadays. They capture hemispherical images of the sky at regular intervals using a fisheye lens. In this paper, we propose a framework for estimating solar irradiance from pictures taken by those imagers. Unlike pyranometers, such sky images contain information about cloud coverage and can be used to derive cloud movement. An accurate estimation of solar irradiance using solely those images is thus a first step towards short-term forecasting of solar energy generation based on cloud movement. We derive and validate our model using pyranometers co-located with our whole sky imagers. We achieve a better performance in estimating solar irradiance and in particular its short-term variations as compared to other related methods using ground-based observations. %Our method shows a significant improvement in estimating strong short-term variations, as compared to methods Hargreaves and Samani~\citep{HSmodel}, Bristow and Campbell~\citep{BCmodel}, Donatelli and Campbell~\citep{DCmodel} and Hunt \emph{et al.}~\citep{Huntmodel}.
\end{abstract}

\setlength{\fboxsep}{0pt}
\setlength{\fboxrule}{0.1pt}

\introduction  %% \introduction[modified heading if necessary]
\label{sec:intro}

Clouds have a significant impact on solar energy generation. They intermittently block the sun and significantly reduce the solar irradiance reaching solar panels. A short-term forecast of solar irradiance is needed for grid operators to mitigate the effects of a drop in power generation. With rapid developments in photogrammetric techniques, ground-based sky cameras are now widely used~\citep{GRSM2016}. These cameras, known as Whole Sky Imagers (WSIs) are upward looking devices that capture images of the sky at regular intervals of time. These images are subsequently used for automatic cloud coverage computation, cloud tracking, and cloud base height estimation. In our research group, we use these imagers to study the effects of clouds on satellite communication links~\citep{dev2018high, Yuan_TGRS15,JSTARS2016}.

Localized and short-term forecasting of cloud movements is an on-going research topic~\citep{shakya2017Solar,jiang2017day,feng2018unsupervised}. Optical flow techniques can be used to forecast images using  anterior frames~\citep{tencon}. Similarly, cloud motion vectors are exploited for solar power prediction from satellite images~\citep{jang2016solar}. Our proposed method for estimating solar irradiance is thus a first step towards solar irradiance forecasting, as the input data used to estimate the irradiance is the same as the one used to forecast the sky condition.

The accurate estimation and prediction of solar energy generation is a challenging task, as clouds greatly impact the total irradiance received on the earth's surface. In the event of clouds covering the sun for a short time, there is a sharp decline of the produced solar energy. Therefore, it is important to model the incoming solar radiation accurately. In this paper, we attempt to measure the rapid fluctuations of solar irradiance  using ground-based sky cameras.

The analysis of clouds and several other atmospheric phenomena is traditionally done using satellite images. However, satellite images have either low temporal or low spatial resolutions. A popular instrument is  Moderate-resolution Imaging Spectroradiometer (MODIS)~\citep{pagano1993moderate}, which is on board the Terra and Aqua satellites and provides a large-scale view of  cloud dynamics and various atmospheric phenomena. Data from MODIS are usually available only twice in a day for a particular location. This is useful for a macro-analysis of cloud formation on the earth's surface. Another illustrative example of such satellite data is the HelioClim-1 database from Global
Earth Observation System of Systems (GEOSS)~\citep{lautenbacher2006global}. It provides hourly and daily average of surface solar radiation received at ground level~\citep{HelioClim2014}. Ouarda et al.\ in~\citep{SEVIRI2016} assessed the solar irradiance from six thermal channels obtained from Spinning Enhanced Visible and Infrared Imager (SEVIRI) instrument. However these are temporal and spatial averages. Solar energy applications requires knowledge of the solar irradiance at specific locations and at all times throughout the day. Therefore, images obtained from satellites are generally not conducive for continuous analysis and prediction, especially in regions where cloud formation is highly localized.

\subsection{Related Work}
Several existing works analyze ground-based images with different meteorological observations. Most of them correlate the cloud coverage obtained from sky images with meteorologists' observations. \citet{Silva2016} validated cloud coverage measurements obtained from ground-based automatic imager and human observations for two meteorological stations in Brazil. \citet{Huo2012} also performed such field experiments for three sites in China. The computation of such cloud coverage percentage is important in solar energy generation. It can hugely impact the amount of solar radiation falling at a particular place. 

The correct estimation of solar irradiance, is particularly important in tropical countries like Singapore, where the amount of received solar irradiance is high. \citet{rizwan2012generalized}  demonstrated that tropical countries are conducive for installing large central power stations powered by solar energy, because of the large amount of incident sunlight throughout the year. Several attempts have been made to estimate the solar radiation from general meteorological measurements via temperature, humidity and precipitation~\citep{HSmodel,DCmodel,BCmodel,Huntmodel}. These existing models aim to provide global solar radiation using different sensors. \citet{alsadi2017estimation} demonstrated such estimation models from the perspective of a photovoltaic (PV) solar field, showing that succeeding rows of PV panels receive less solar radiation than that of first row. They also provided an analytical solution by including the design parameters in the estimation model.

In addition to solar irradiance estimation, there have been several efforts in forecasting the solar irradiance, with a lead time of few minutes. \citet{baharin2016short} proposed a machine-learning forecast model for PV power output, using Malaysia as the case study. Similarly \citet{chu2015short} used a reforecasting method to improve the PV power output forecasts with a lead time of $5$, $10$, and $15$ minutes.
Satellite images have also been used in the realm of solar analytics. \citet{mueller2004rethinking} proposed a clear sky model that is based on radiative transfer models obtained from Meteosat's atmospheric parameters. However, satellite data have lower temporal and spatial resolutions. 

Recently, with the development of low-cost photogrammetric techniques, sky camera are being deployed for such purposes. These sky cameras have both high temporal and high spatial resolutions, and are able to provide a more localized information about the atmospheric events. \citet{alonso2015} used sky cameras to quantify the total solar radiation. \citet{yang2015expanding} studied these solar irradiance variability using entropy and covariance. \citet{dev2018solar} used triple exponential smoothing for analyzing the seasonality of the solar irradiance. However, these approaches do not model the sharp short-term variations of solar radiation.

\subsection{Outline}
In this paper, we use images obtained from WSIs to accurately model the fluctuations of the solar radiation. There are several advantages of using a WSI for this instead of a pyranometer. Common weather stations generally use a solar sensor that measures the total solar irradiance. It is a point measurement providing scalar information for a particular location and does not provide information on clouds and their evolution over time. On the other hand, the wide-angle view of a ground-based sky camera provides us extensive information about the sky. It allows for the tracking of clouds over successive image frames, and also to predict their future location. In this paper, we attempt to model solar irradiance from sky images. This can also help in solar energy forecasting, which is useful in photovoltaic systems~\citep{solar_PV}.

The main contributions of this paper are as follows:
\begin{itemize}
\item We develop a framework to accurately estimate and track the rapid fluctuations of solar irradiance; 
\item We propose a method for estimating solar irradiance using ground-based sky camera images;
\item We conduct extensive benchmarking of our proposed method with other solar irradiance estimation models.
\end{itemize}

The rest of the paper is organized as follows. Section~\ref{sec:data} describes our experimental setup that captures the sky/cloud images and collects other meteorological sensor data. Our framework for estimating solar irradiance is presented in Section~\ref{sec:solarmodel}. Section~\ref{sec:comparemodels} discusses the evaluation of our approach, and its benchmarking with other existing solar estimation models. We discuss the possible applications of our approach in Section~\ref{sec:discuss}. We also point out a few limitations of our approach, and ways to address them. Section~\ref{sec:conclusion} concludes the paper.

\section{Data Collection}
\label{sec:data}
Our experimental setup consists of weather stations and ground-based WSIs. These devices are co-located on the rooftop of our university building in Singapore ($1.34^{\circ}$N, $103.68^{\circ}$E). They continuously capture  various meteorological data, and we archive them for subsequent analysis. 

\subsection{Whole Sky Imager (WSI)}
Commercial WSIs are available in the market. However, those imagers have high cost, low image resolution, and little flexibility in operation. In our research group, we have designed and built custom low-cost high-resolution sky imagers, which we call WAHRSIS, i.e.\ Wide Angle High Resolution Sky Imaging System~\citep{WAHRSIS}. A WAHRSIS imager essentially consists of a high-resolution digital single-lens reflex (DSLR) camera with a fish-eye lens and an on-board micro-computer. %The DSLR camera has a digital imaging sensor, instead of the traditional photographic film. 
The entire device is contained inside a weather-proof box with a transparent dome for the camera. Over the years, we have built several versions of WAHRSIS~\citep{WAHRSIS,IGARSS2015a}. They are now deployed at several locations around our university campus, capturing images of the sky at regular intervals. 

\begin{figure}[htb]
\begin{center}
\includegraphics[width=0.7\textwidth]{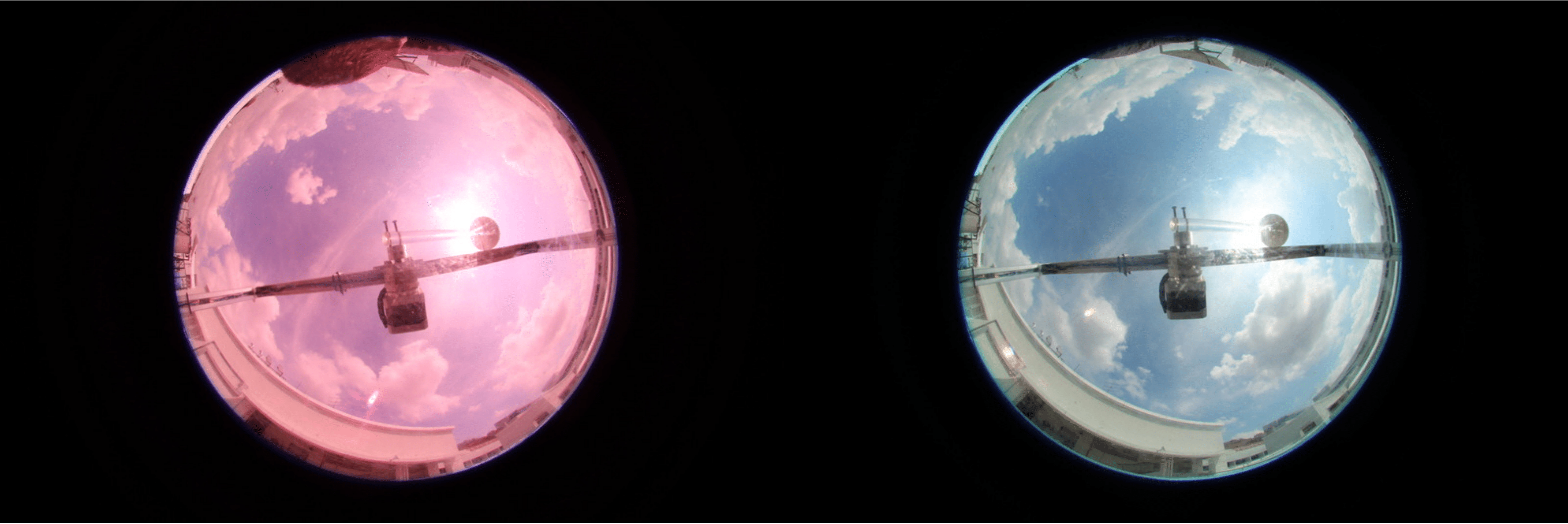}\\
\makebox[0.35\textwidth][c]{\small (a) With auto white-balancing}
\makebox[0.35\textwidth][c]{\small (b) With custom white-balancing}
\caption{We use custom white-balancing for correcting the white balance.
\label{fig:white-balance}}
\end{center}
\end{figure}

Our WAHRSIS camera  is calibrated with respect to white balancing, geometric distortions and vignetting. The imaging system in WAHRSIS is modified so that it captures the near-infrared region of the spectrum. Hence, the red channel of the captured image is more prone to saturation, which renders the captured image reddish in nature. Therefore, we employ custom white balancing in the camera, such that it compensates the alteration owing to the near-infrared capture. Figure~\ref{fig:white-balance} depicts the captured images obtained from automatic and custom white balancing.

\begin{figure}[htb]
\begin{center}
\includegraphics[width=0.7\textwidth]{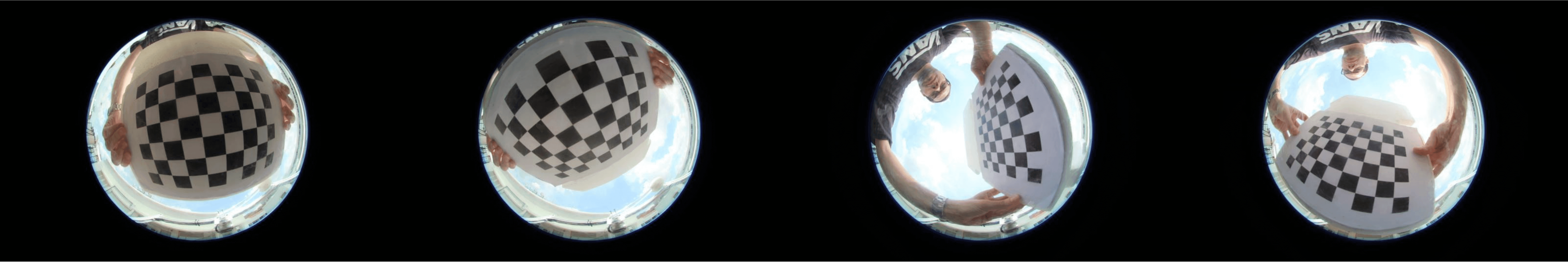}
\caption{We position the checkerboard at various locations for  geometric camera calibration. 
\label{fig:calibration_images}}
\end{center}
\end{figure}

We use the popular toolbox by \citet{scaramuzza2006toolbox} for the geometric calibration of WAHRSIS. This process involves the computation of the intrinsic parameters of the camera. We use a black-and-white regular checkerboard pattern, and position it at various points around the  camera. Figure~\ref{fig:calibration_images} illustrates a few sample positions of the checkerboard in the calibration process. Using  the corner points and the known dimensions of the pattern, we can estimate the intrinsic parameters of the camera.

\begin{figure}[htb]
\begin{center}
\includegraphics[width=0.45\textwidth]{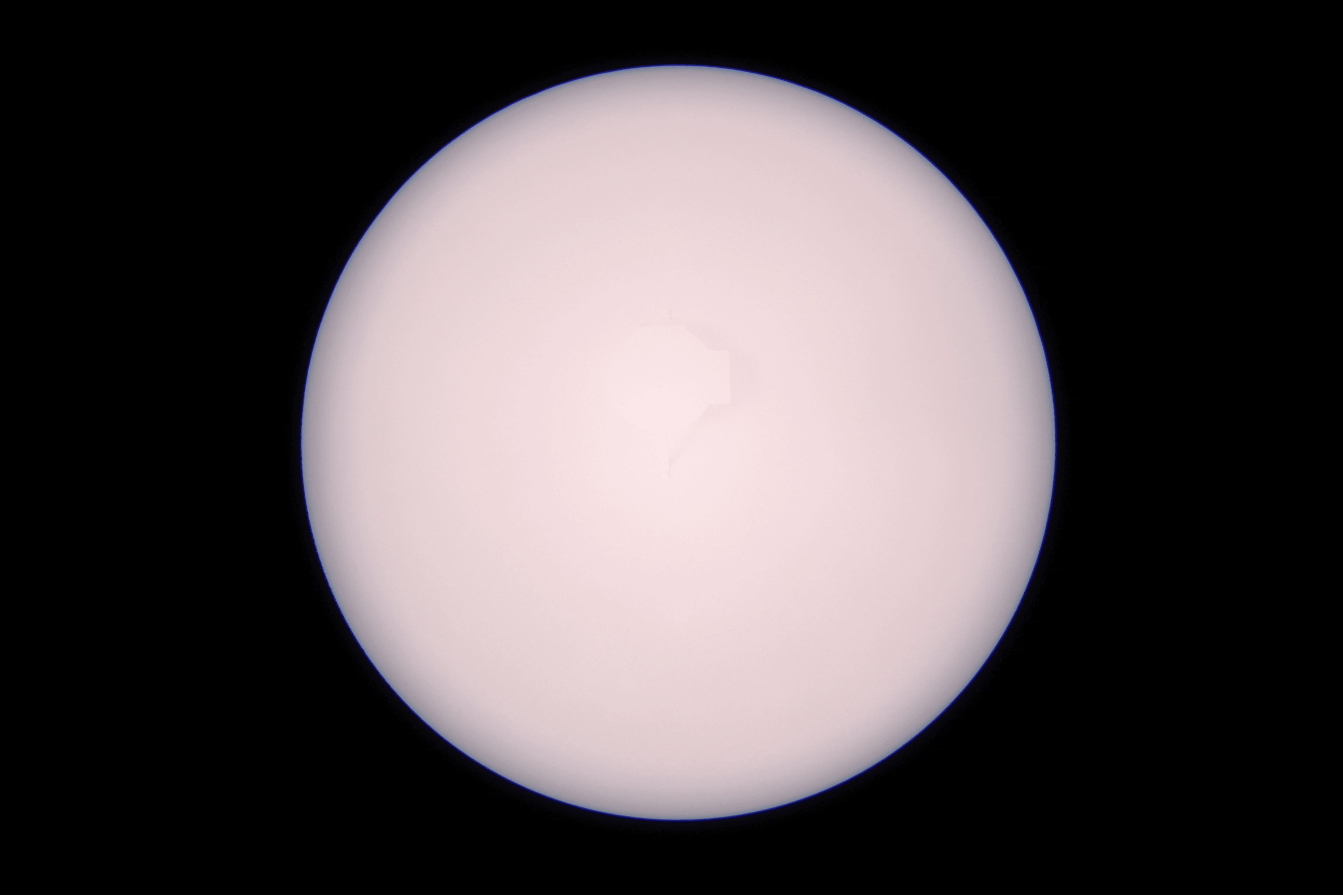}
\caption{Reference image captured inside the uniformly-illuminated integrating sphere.
\label{fig:sphere}}
\end{center}
\end{figure}

Finally, we apply vignetting correction to the images captured by our sky camera. Owing to the fish-eye nature of the lens, the area around the centre of the lens is brighter than at the sides. We use an integrating sphere to correct this variation of illumination. Figure~\ref{fig:sphere} depicts an image captured inside an integrating sphere that provides a uniform illumination distribution in all directions. We use luminance characteristics from this reference image to correct the  images captured by our sky camera.

\subsection{Weather Station}
In addition to the sky imagers, we have also installed co-located weather stations. We use \emph{Davis Instruments 7440 Weather Vantage Pro} for our recordings. It measures rainfall, total solar radiation, temperature and pressure at intervals of $1$ minute. The resolution of the tipping-bucket rain gauge is $0.2$ mm/tip.

\begin{figure}[htb]
\begin{center}
\includegraphics[width=0.5\textwidth]{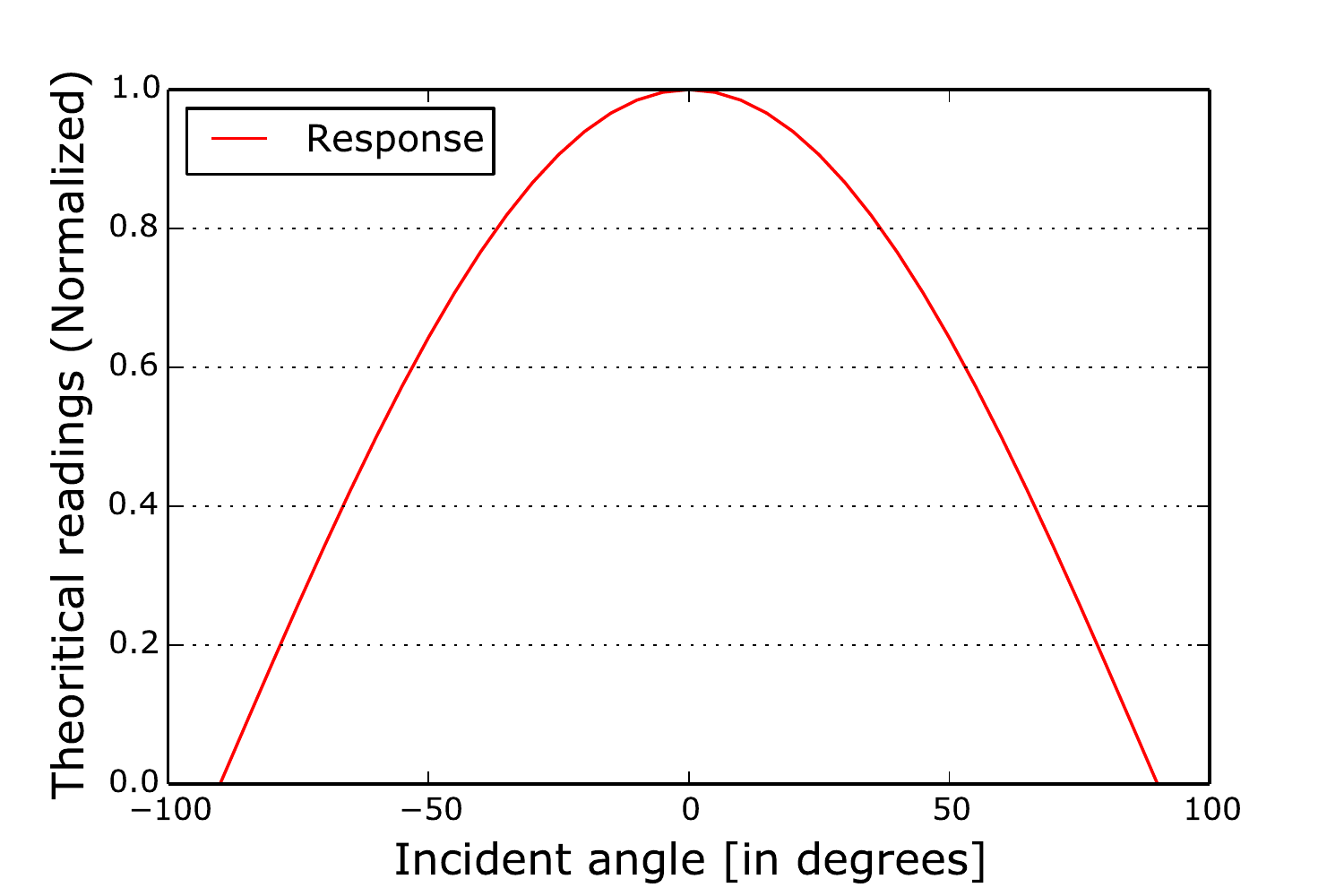}
\caption{Response of the solar sensor on a clear day with varying solar incident angle.
\label{fig:sensor-reading}}
\end{center}
\end{figure}

It also includes a solar pyranometer measuring the total solar irradiance flux density in W/$\mbox{m}^2$. This consists of both direct and diffuse solar irradiance components. The solar sensor integrates the solar irradiance across all angles, and provides the total solar irradiance. On a clear day with no occluding clouds, the solar sensor can be approximated by a typical cosine response, shown in Figure~\ref{fig:sensor-reading} for varying degrees of solar incident angle. The solar sensor reading is highest around noon when the incident angle of sun rays is at the minimum, whilst the reading is low during morning and evening hours.

The solar radiation under a clear sky can be modeled using the solar zenith angle and the earth's eccentricity. Several clear sky models have been developed for various regions. The best clear-sky model for Singapore is provided by \citet{dazhi2012estimation}. We performed a comparison of various clear sky models in Singapore~\citep{dev2017study}, and found that the \citet{dazhi2012estimation} model provides a good estimate of the clear sky irradiance. It models clear-sky Global Horizontal Irradiance (GHI) $G_c$   as follows: 
\begin{align}
%\label{eq:GHI-model}
G_c = 0.8277E_{0}I_{sc}(\cos\alpha)^{1.3644}e^{-0.0013\times(90-\alpha)},
\end{align}
where $E_{0}$ is the eccentricity correction factor for earth, $I_{sc}$ is the solar irradiance constant ($1366.1$Watt/$\mbox{m}^2$), and $\alpha$ is the solar zenith angle (measured in degrees). The factor $E_{0}$ is calculated as:
\begin{equation*}
\begin{aligned}
\label{eq:E0value}
E_0 = 1.00011 + 0.034221\cos(\Gamma) + 0.001280\sin(\Gamma) + 0.000719\cos(2\Gamma) + 0.000077\sin(2\Gamma),
\end{aligned}
\end{equation*}
where $\Gamma = 2\pi(d_n-1)/365$ is the day angle (measured in radians) and $d_n$ is the day number of the year. 

\begin{figure}[htb]
\begin{center}
\includegraphics[width=1\textwidth]{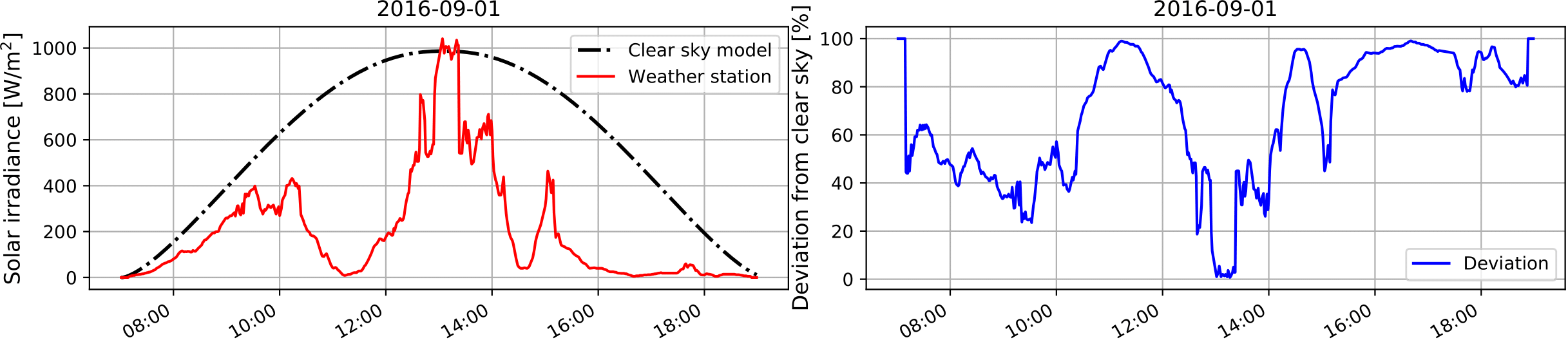}\\
\makebox[0.45\textwidth][c]{(a) \small Measured solar irradiance along with clear-sky model.}
\makebox[0.45\textwidth][c]{(b) \small Percentage deviation of solar irradiance from clear sky data.}
\caption{Solar irradiance measurements on the 1st of September 2016. Note the rapid fluctuations of high magnitude in the measurements.
\label{fig:Cos_response}}
\end{center}
\end{figure}

As an illustration, we show the clear-sky radiation for the 1st of September 2016 in Figure~\ref{fig:Cos_response}, compared to the actual solar irradiance measured by our weather station. We also show the deviation of the measured solar radiation from the clear-sky model. We observe that there are rapid fluctuations in the measured readings. In our previous work~\citep{IGARSS16_solar}, we observed that these rapid fluctuations are caused by the incoming clouds that obstruct the sun from direct view. Such information about the cloud profile and its formation cannot be obtained from a point-source solar recording. Therefore, we aim to model these rapid fluctuations in the measured solar radiation from wide-angle images captured by our sky cameras.

\section{Modeling Solar Irradiance}
\label{sec:solarmodel}
This section presents our model for computing solar irradiance from images captured by a whole sky imager. We sample pixels using a cosine weighted hemispheric sampling to simulate the behavior of a pyranometer based on the fisheye camera lens. We then compute the relative luminance using the image capturing parameters after gamma correction. We finally derive an empirical fitting function to scale the computed luminance estimates to match measured irradiance values.

\subsection{Cosine Weighted Hemispheric Sampling}
\begin{figure}[htb]
\begin{center}
\includegraphics[width=0.9\textwidth]{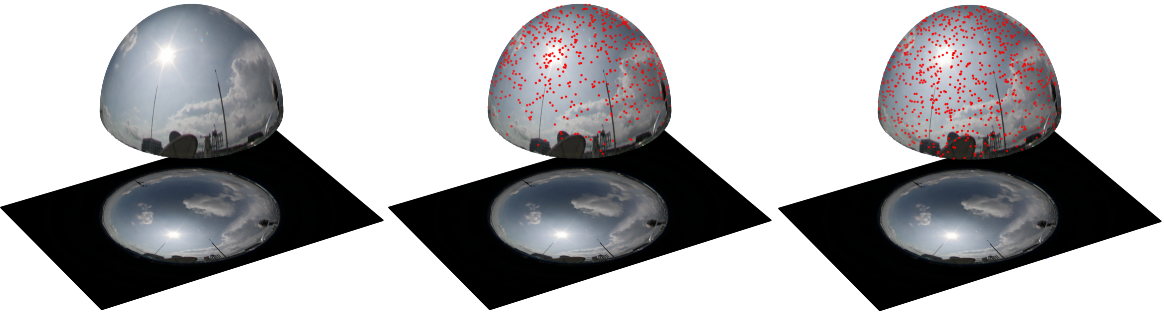}\\
\parbox{0.3\textwidth}{(a) Projection of the original image on a hemisphere.}
\parbox{0.3\textwidth}{(b) Cosine hemispheric sampling with origin at the top.}
\parbox{0.3\textwidth}{(c) Applying a rotation matrix to center at the sun location.}\\ 
\vspace{0.3cm}
\includegraphics[width=0.9\textwidth]{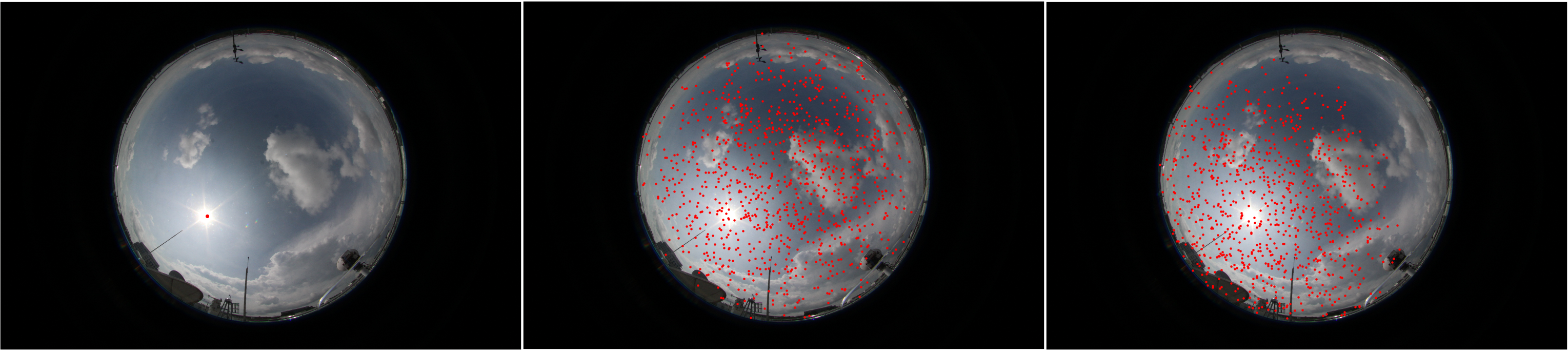}\\
\parbox{0.3\textwidth}{(d) Original image with detected sun location in red.}
\parbox{0.3\textwidth}{(e) Projection of the sampled points onto the image.}
\parbox{0.3\textwidth}{(f) Projection of the rotated sampled points onto the image.}
\caption{Cosine weighted hemispheric sampling process.
\label{fig:sampling}}
\end{center}
\end{figure}

The behavior of our fisheye lens with focal length $f$ is modeled by the equisolid equation $r=2f \sin(\theta/2)$, relating the distance ($r$) of any pixel from the center of the image to its incident light ray elevation angle ($\theta$). This allows to project a captured image onto the unit hemisphere, as shown in Figure~\ref{fig:sampling}. 

The solar irradiance is composed of a direct component relating the sun light reaching the earth without interference, as well as diffuse and reflected components. Given the high resolution of our images, we consider randomly sampled pixel locations on the hemisphere as input to the luminance computation. We follow a cosine weighted hemispheric distribution function, the center of which is at the location of the sun. This is because clouds in the circumsolar region have the highest impact on the total solar irradiance received on the earth's surface~\citep{IGARSS16_solar}. We provide more emphasis to the clouds around the sun, as compared to those near the horizon. In our previous work~\citep{dev2016estimation}, we used a cloud mask around the sun to estimate the solar irradiance. However, this requires the additional step of optimizing the size of the cropped image for best results. Therefore, we adopt the strategy of cosine weighted hemispheric sampling. 

The first step is to compute the sampled locations from the top of the unit hemispheric dome. Each of the locations are computed as follows, using two random floating points $R_1$ and $R_2$ as input, where $(0 \leq R_1, R_2 \leq 1)$: 
\[\phi = 2\pi R_1,\ \theta = \arccos(\sqrt{R_2})\]
\begin{equation}
     \begin{bmatrix}
         x \\
         y \\
         z
        \end{bmatrix}=\begin{bmatrix}
         \sin(\theta)\cdot \cos(\phi) \\
         \sin(\theta)\cdot \sin(\phi) \\
         \cos(\theta)
        \end{bmatrix}
  \end{equation}
This is represented in Figure~\ref{fig:sampling}. 

The second step is to detect the location of the sun using a thresholding method. This is needed to align the center of the previously computed distribution (i.e.\ top of the hemispheric dome) to the actual sun location in the unit sphere. We choose a threshold of $240$ in the red channel $R$ of the $RGB$ captured image, and compute the centroid of the largest area above the threshold~\citep{IGARSS16_calib}. We then compute the rotation matrix transforming the z-axis unit vector to the unit vector pointing towards the sky. We apply this rotation to all the sampled points, resulting in Figure~\ref{fig:sampling}.  This means that the number of sampled points in a region of the hemisphere is proportional to the cosine of the angle between the sun direction and the direction to that region. We experimentally concluded that this achieves a good balance between all irradiance components. We consider the pixel values of a total of $5000$ points sampled using this method as input for the irradiance estimation.

\subsection{Relative Luminance Calculation}

For each of the $i$ sampled pixels in the $RGB$ image, we compute its luminance value using the following formula. The formula is proposed in SMPTE Recommended Practice 177~\citep{smpte1993rp} to compute the luminance of an image from the \emph{R}, \emph{G} and \emph{B} values of the \emph{RGB} image. 
\[Y_i = 0.2126\cdot R_i  + 0.7152\cdot G_i + 0.0722\cdot B_i\]

The JPEG compression format encodes images after applying a gamma correction. This non-linearity mimics the behavior of the human eye. This needs to be reversed in order to compute the irradiance. We use a gamma correction factor of $2.2$, which is most commonly used in imaging systems~\citep{Poynton03}. We thus apply the following formula, assuming pixel values normalized between $0$ and $255$:
\[Y_i' = 255{(Y_i/255)}^{2.2}\]

We then average the pixel values across all the $i$ sampled points in the image, and denote it by $\mathcal{N} = (1/n)\sum_{i=1}^{n} Y_i'$, the average luminance value of the sampled points from the image. 

However, each image of the sky camera is captured with varying camera parameters such as ISO, F-number and shutter speed. These camera parameters can be read from the image metadata, and are useful to estimate the scene luminance. The amount of brightness of the sampled points $\mathcal{N}$, is proportional to the number of photons hitting the camera sensor. This relationship between scene luminance and pixel brightness is linear~\citep{hiscocks2011measuring}, and can be modeled using the camera parameters:
\[\mathcal{N} = K_c \left( \frac{e_t\cdot S}{f_s^2}\right) \mathcal{L}_s\]
where $\mathcal{N}$ is the pixel value, $K_c$ is a calibration constant, $e_t$ the exposure time in seconds, $f_s$ the aperture number, $S$ the ISO sensitivity and $\mathcal{L}_s$ the luminance of the scene.

We can thus compute the relative luminance $\mathcal{L}_r$ as follows:
\[\mathcal{L}_r = \mathcal{N} \left( \frac{f_s^2}{e_t\cdot S}\right)\]

\subsection{Modeling Irradiance from Luminance Values}
Using our hemispheric sampling and relative luminance computation, we therefore have one relative luminance value $\mathcal{L}_r$ per image. We use this relative luminance value to estimate the solar radiation. The usual sunrise time in Singapore is between $6$:$40$am and $7$:$05$am, and sunset time is approximately between $6$:$50$pm and $7$:$10$pm local time. 
This information is obtained from \citep{nea-weather}. Therefore, we consider images captured in the time interval of $7$:$00$am till $7$:$00$pm. 

We use our ground-based whole sky images captured during the period from January $2016$ till August $2016$ to model the solar radiation. The solar irradiance is computed as the flux of radiant energy per unit area normal to the direction of flow. The first step in estimating irradiance from the luminance is thus to cosine weight it according to its direction of flow. We weight our measurements according to the solar zenith angle $\alpha$. This is based on empirical evidences of our experiments on solar irradiance estimation. The modeled luminance $\mathcal{L}$ is expressed as: 
\begin{equation*}
%\label{eq:GHI-model}
\mathcal{L} = \mathcal{L}_r(\cos\alpha)
\end{equation*}

\begin{figure}[htb]
\begin{center}
\includegraphics[width=0.9\textwidth]{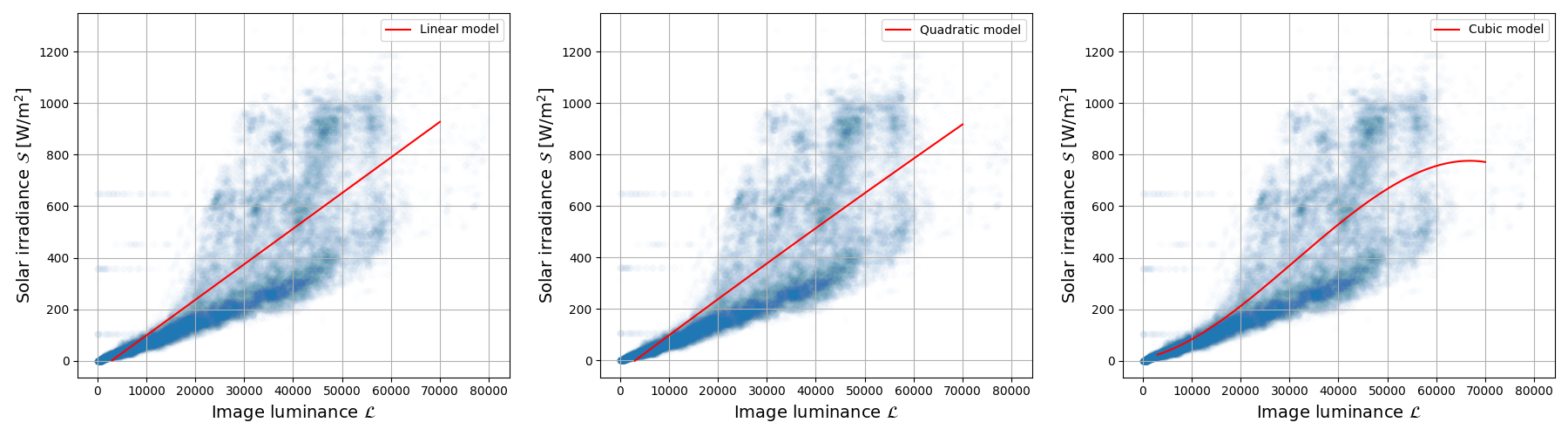}\\
\parbox{0.3\textwidth}{\centering (a) Linear model.}
\parbox{0.3\textwidth}{\centering (b) Quadratic model.}
\parbox{0.3\textwidth}{\centering (c) Cubic model.}
\caption{Empirical fit between solar irradiance and image luminance computed with our proposed framework. We observe that it deviates from linearity at higher luminance values. Higher order polynomials are ill-conditioned. 
\label{fig:solar_model}}
\end{center}
\end{figure}

Let us assume that the actual solar radiation recorded by the weather station is $\mathcal{S}$. We check the nearest weather station measurement for all the images captured by WAHRSIS between April $2016$ till December $2016$. Figure~\ref{fig:solar_model} shows the scatter plot between the image luminance and solar radiation. The majority of the data follows a linear relationship between the two. However, it deviates from linearity for higher values of luminance. This is mainly because of the fact that the mapping between scene luminance and obtained pixel value in the camera sensor becomes non-linear for large luminances. A more detailed discussion on this is provided in Section~\ref{sec:discuss}.

We attempt to fit a linear model and other higher-order polynomial regressors to model the relationship between image luminance from sky camera images and the measured solar radiation. Figure~\ref{fig:solar_model} shows the best fit curve for several orders of polynomial function. In order to provide an objective evaluation of the different models, we also compute the RMSE value between the actual and regressed values. Table~\ref{tab:model-fitting} summarises the performance of the different order polynomials. We observe that lower order polynomials of degree $1$ and $2$ perform slightly inferior to those of higher order polynomials. %We observe that the performance of cubic, quartic and quintic models are similar. 
However, higher-order polynomial models are ill-conditioned. Therefore, we choose the cubic model as our proposed model to model the measured solar radiation $\mathcal{S}$ from the image luminance $\mathcal{L}$. This is based on the assumption that the mapping from scene luminance to pixel values in the captured image is linear for lower luminance values, and it behaves in a non-linear fashion for higher luminance values. We use this selected model in all our subsequent discussions and evaluations.

\begin{table}[htb]
\normalsize
\centering
\caption{Performance evaluation of various polynomial order regressors. We measure the RMSE value for each of the models.}
\label{tab:model-fitting}
\begin{tabular}{rc}
\hline
\textbf{Proposed models} & \textbf{RMSE} (W/$\mbox{m}^2$) \\
\hline 
Linear (degree $1$) & 178.27 \\
Quadratic (degree $2$) & 178.26 \\
Cubic  (degree $3$) & 176.57 \\
Quartic (degree $4$) & 176.52 \\
Quintic (degree $5$) & 176.49 \\
\hline
\end{tabular}
\end{table}

We model solar radiation as: $\mathcal{S} = a_3\times\mathcal{L}^3 + a_2\times\mathcal{L}^2 + a_1\times\mathcal{L}+ a_0$, with $a_3=-4.25e-12$, $a_2=3.96e-07$, $a_1=0.00397$ and $a_0=7.954$ for our data. %Therefore, our proposed methodology in estimating solar irradiance from luminance is:
%\begin{align}
%\label{eq:propmodel}
%\mathcal{S} = (-4.25e-12)\mathcal{L}^3 + (3.96e-07)\mathcal{L}^2 + (0.00397)\times\mathcal{L}+ 7.954.
%\end{align}
This model is derived specifically for the equatorial region like Singapore, and the regression constants are based on our WAHRSIS sky imaging system. They would have to be fine-tuned for other regions and different imaging systems using our methodology. To facilitate this, we make the source code of all the simulations in this paper available online at \url{https://github.com/Soumyabrata/estimate-solar-irradiance}.

\section{Experimental Validation}
\label{sec:comparemodels}
In this section, we evaluate the accuracy of our proposed approach. It is derived based on WAHRSIS images captured from January to August $2016$. We also use these images to evaluate the accuracy of our proposed model. Furthermore, we benchmark our algorithm with other existing solar radiation estimation models.

\subsection{Evaluation}

\begin{figure*}[htb]
\centering
\includegraphics[width=0.95\textwidth]{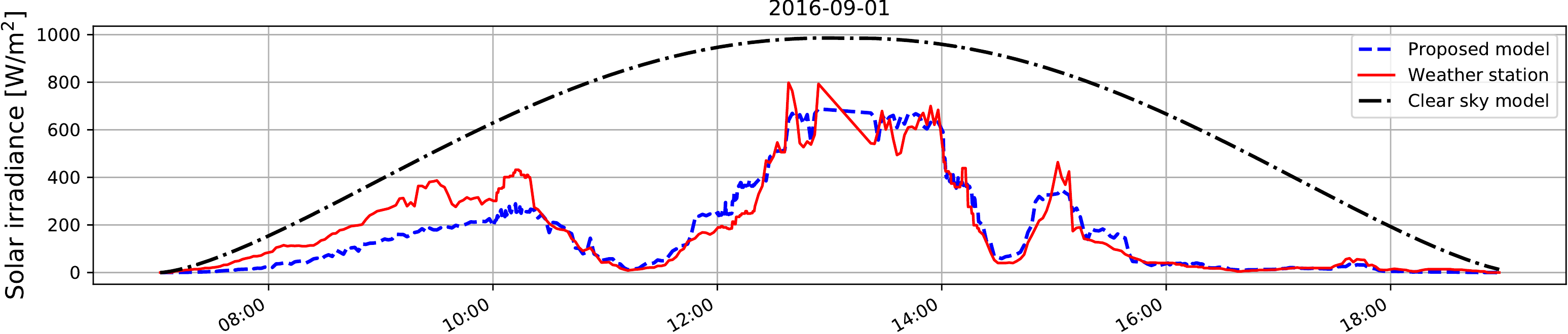}
\caption{Measured weather station data (in red) vs.\  clear sky radiation (in black) as on 01-Sep-2016. The sampling interval between two measurements is $2$ minutes.
\label{fig:tracksun}}
\end{figure*}

One of the main advantages of our approach is that all rapid fluctuations of solar radiation can be accurately tracked from the image luminance. We illustrate this by providing the measured solar readings of 01-Sep-2016 in Figure~\ref{fig:tracksun}. The clear-sky model follows a cosine response and is shown in black, while the measured solar recordings are shown in red. We normalize our computed luminance in such a manner that it matches the measured solar readings. We multiply each data point with a conversion factor, such that the distance between corresponding inter-samples of luminance and weather station recordings is minimized (cf.\ Appendix A for details). We use this normalization factor in order to map the computed image luminance to have a  similar scale to the cosine clear sky model. We observe that our computed luminance from the whole-sky image and the measured solar radiation closely follow each other. We emphasize here that it is  important to accurately track the rapid solar fluctuations. Unlike other solar estimation models based on meteorological sensor data, our proposed model can successfully estimate the \emph{peaks} and \emph{troughs} of solar readings. 

\begin{figure}[htb]
\begin{center}
\includegraphics[width=0.45\textwidth]{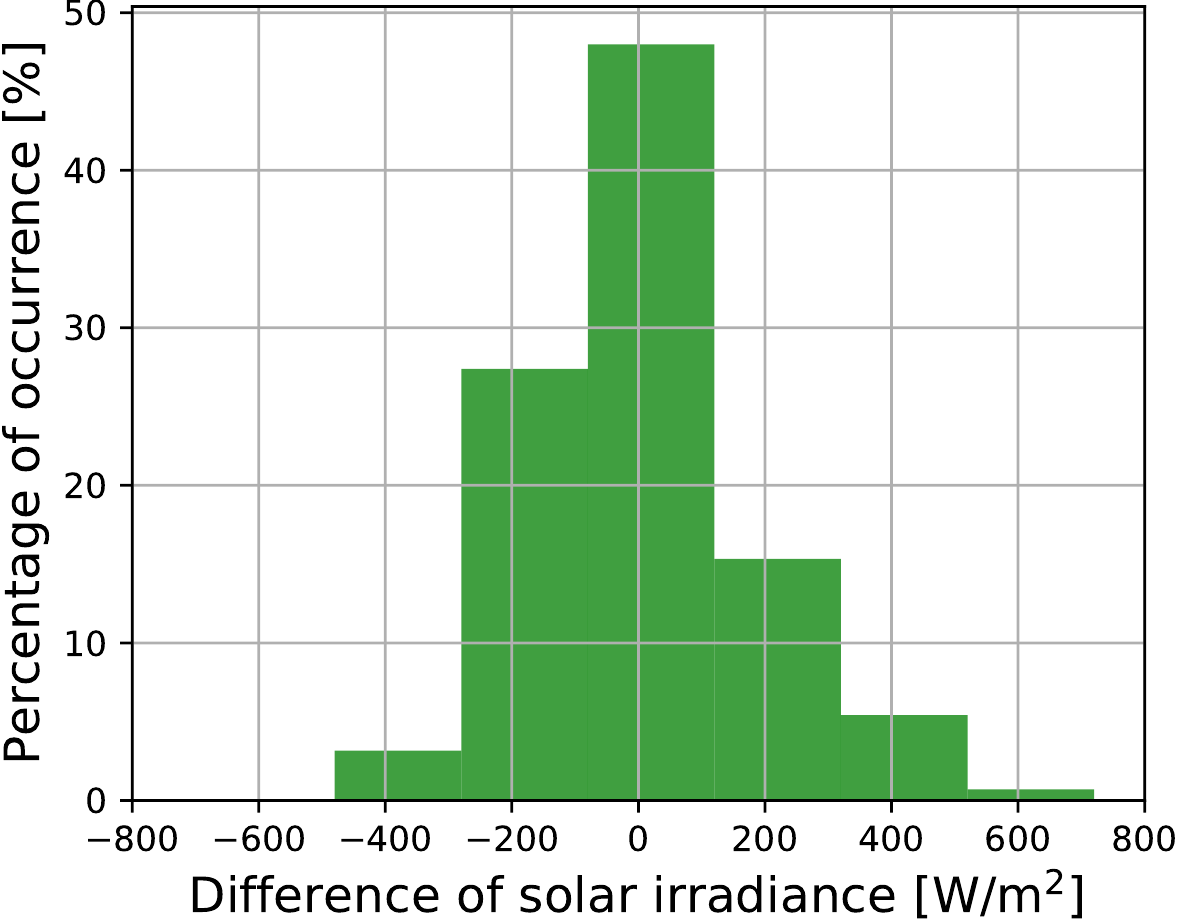}
\caption{Histogram of differences between estimated and actual solar irradiance. 
\label{fig:HOD}}
\end{center}
\end{figure}

Using our proposed methodology, we compute the luminance of all the captured images. Subsequently, using our proposed cubic model, we estimate the corresponding solar radiation values. The estimated solar irradiance values are compared with the actual irradiance values obtained from the solar sensors in the co-located weather station, which serve as the ground-truth measurements. Figure~\ref{fig:HOD} shows the histogram of differences between the estimated and actual solar radiation. We observe that the estimates do not deviate much from the actual solar radiation. Nearly half ($47.9\%$) of data points are concentrated in the range [$-100$,$+100$] W/$\mbox{m}^2$.

\subsection{Benchmarking}

\begin{figure*}[htb]
\centering
\includegraphics[width=0.95\textwidth]{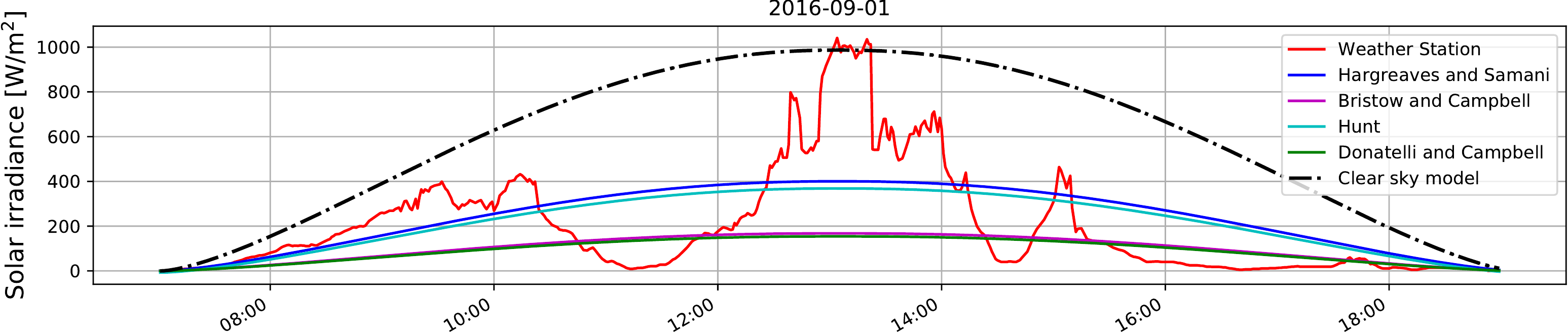}
\caption{Comparison amongst different benchmarking solar estimation models, along with clear sky model and measured solar irradiance on 01-Sep-2016. Most of the existing algorithms fail to capture the rapid fluctuations of the measured solar irradiance. 
\label{fig:othermodels}}
\end{figure*}

We benchmark our proposed approach with other existing solar estimation models. To the best of our knowledge, there are no proposed models to estimate short-term fluctuations of solar irradiance from ground-based images. However, most remote sensing analysts have been using other meteorological sensor data, e.g.\ temperature, humidity, rainfall and dew point temperature to estimate daily solar irradiance. One of the pioneer works was done by \citet{HSmodel}, who proposed a model based on daily temperature variations. \citet{DCmodel} improved the model by including clear sky transitivity as one of the factors. On the other hand, \citet{BCmodel} also proposed a new model of solar radiation estimation, by including the atmospheric transmission coefficient. Subsequently, \citet{Huntmodel} showed that the solar estimation model can be further improved by incorporating   precipitation data in the model. We benchmark our proposed approach with these different existing models. We illustrate the various benchmarking models in Figure~\ref{fig:othermodels}. Unfortunately, most of these other approaches fail to capture the short-term variations of the  solar radiation. 

We calculate the Root Mean Square Error (RMSE) of the estimated solar radiation and Spearman's rank correlation coefficient as evaluation metrics. The RMSE of an estimation algorithm represents the standard deviation of the actual and estimated solar radiation values. Spearman correlation is a non-parametric measure to characterize the relationship between measured and estimated solar radiation, which does not assume that the underlying dataset are derived from a normal distribution. We report both metrics in Table~\ref{tab:corr_results}. Our proposed approach achieves the best results amongst all methods. Note that the training and testing set of images are identical, and all images are considered for benchmarking purposes.

\begin{table}[htb]
\normalsize
\centering
\caption{Benchmarking of our proposed approach with other solar radiation estimation models.} %All correlation values have p-value equal to $0$.}
\label{tab:corr_results}
\begin{tabular}{rcc}
\hline
\textbf{Methods} & \textbf{RMSE} (W/$\mbox{m}^2$) & \textbf{Correlation}\\
\hline 
Proposed approach & \textbf{178.27} & \textbf{0.86}\\
\citet{HSmodel} & 982.35 & 0.67\\
\citet{DCmodel} & 324.48 & 0.67\\
\citet{BCmodel} & 318.07 & 0.68\\
\citet{Huntmodel} & 922.66  & 0.65\\
\hline
\end{tabular}
\end{table}

Furthermore, we check if our proposed model generalizes well with random samples of our captured sky camera images. We choose a random selection of images as the training set, and fit our linear regressor on these selected training images. The RMSE values are then calculated on these training images. We perform this analysis for varying percentages of training images. Each experiment is performed $100$ times to remove any selection bias.

\begin{figure}[htb]
\begin{center}
\includegraphics[width=0.9\textwidth]{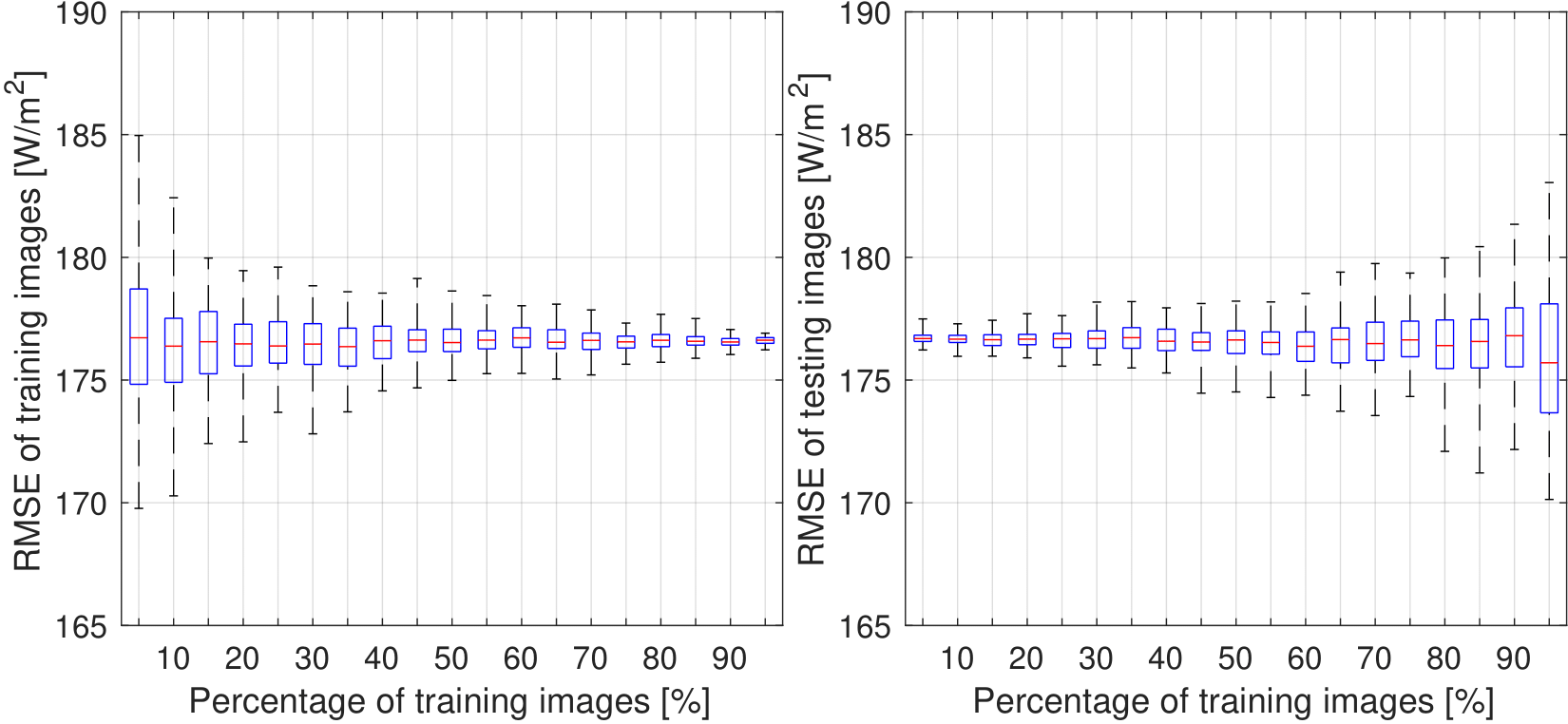}\\
\parbox{0.45\textwidth}{\centering (a) Performance on training set.}
\parbox{0.45\textwidth}{\centering (b) Performance on test set.}
\caption{Effect of the percentage of training images on RMSE values. The lower and upper end of each box represents the $25^{th}$ and $75^{th}$ percentiles of the data, and the red line represents the median. Each experiment is conducted $100$ times with a random choice of training and test sets.
\label{fig:train_test}}
\end{center}
\end{figure}

Figure~\ref{fig:train_test}(a) shows the results on training images. We observe that the variation of the RMSE values gradually decreases as we increase the number of training images. Moreover, we check the variation of RMSE values when the test images are not identical as training images. Once we choose a random selection of images as training set, the remaining images are considered as the test set. We show the RMSE on such images in Fig~\ref{fig:train_test}(b). As expected, the variation of RMSE values increases with higher percentage of training images. The linear regressor model overfits the data, and provides higher variation in the error when tested on a fewer test images. However, the average RMSE does not vary much in all cases. Therefore we conclude that our proposed model is free from selection bias, and generalizes well with random selection of training and testing images.

\begin{figure}[htb]
\begin{center}
\includegraphics[width=0.9\textwidth]{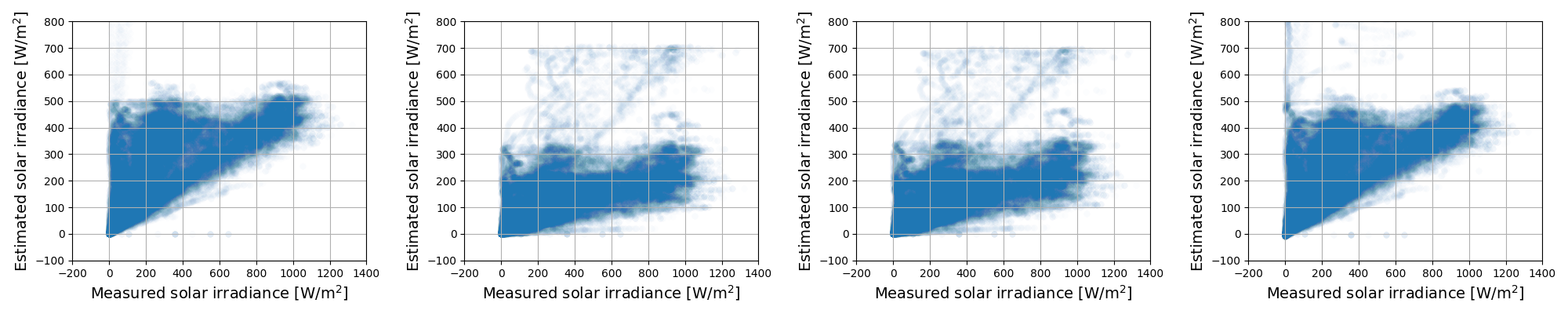}\\
\parbox{0.22\textwidth}{\centering (a)\small \citet{HSmodel}.}
\parbox{0.22\textwidth}{\centering (b)\small \citet{DCmodel}.}
\parbox{0.22\textwidth}{\centering (c)\small \citet{BCmodel}.}
\parbox{0.22\textwidth}{\centering (d)\small \citet{Huntmodel}.}
\caption{Scatter plot between measured solar irradiance and estimated solar irradiance for the benchmarking algorithms.
\label{fig:scatter-other}}
\end{center}
\end{figure}

We show the scatter plot between the measured solar radiation and estimated solar radiation for the different benchmarking algorithms in Figure~\ref{fig:scatter-other}. We observe that there is no strong correlation for most of these existing algorithms. This is because meteorological sensor data alone, with no cloud information cannot determine the sharp fluctuations of the solar radiation. 
This is an important limitation of these models, which we have attempted to address in this paper. Our model based on sky images has additional information about cloud movement and its evolution, which is the fundamental reason behind rapid solar radiation fluctuations. In our proposed model, most of these short-term variations are captured (cf.\ Figure~\ref{fig:tracksun}).

\section{Discussion}
\label{sec:discuss}

\subsection{Short-term Forecasts}
\begin{figure}[htb]
\begin{center}
\includegraphics[width=0.9\textwidth]{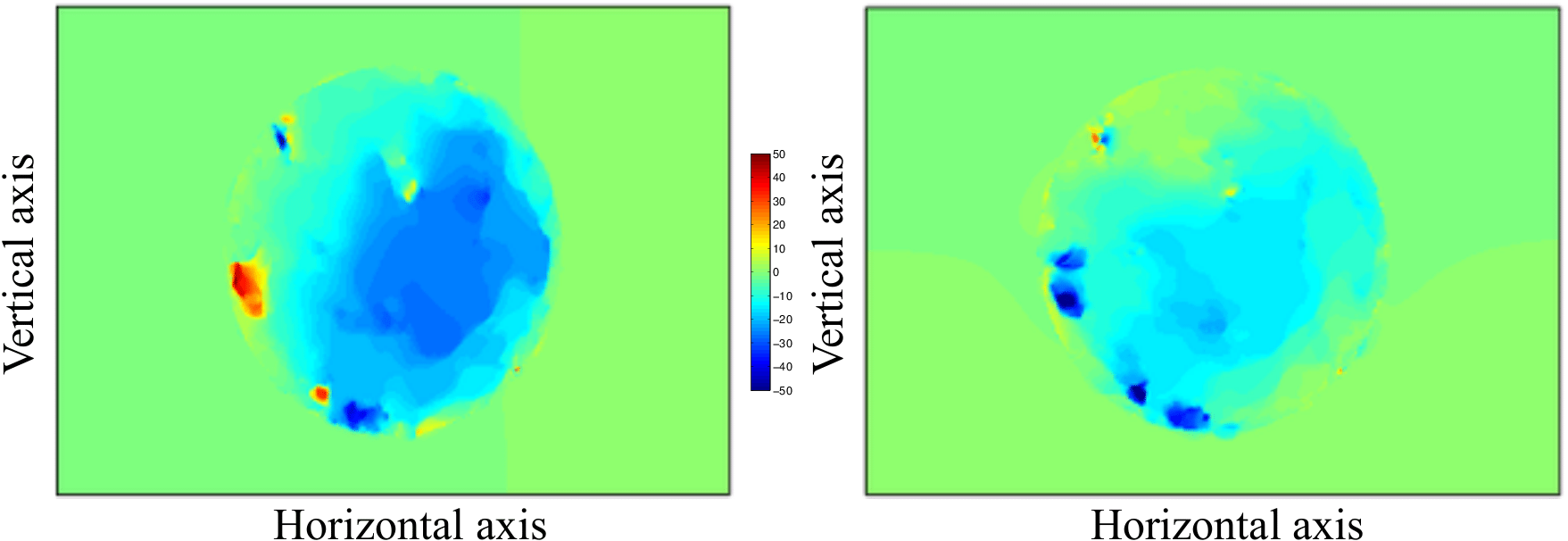}\\
\parbox{0.4\textwidth}{\centering \small (a) Horizontal translation.}
\parbox{0.4\textwidth}{\centering \small (b) Vertical translation.}
\caption{Horizontal and vertical translation of pixels between two successive frames, computed using optical flow. 
\label{fig:vectorflow}}
\end{center}
\end{figure}

Our proposed approach can estimate the solar radiation accurately with the smallest RMSE compared to other models. The main advantage of our approach is that it can be used on predicted images as well, opening the potential for short term solar irradiance forecasting, which is needed in the solar energy field. As an initial case study, we have exploited optical flow techniques to estimate the direction and flow of cloud motion vectors between two successive image frames. We use the $(B-R)/(B+R)$ ratio channel of the sky/cloud image, where $B$ and $R$ are the blue and red channels respectively. We implement an optical flow technique~\citep{flow-pgm} that uses a simpler conjugate gradient solver to obtain the flow field. Figure~\ref{fig:vectorflow} illustrates the estimated flow field.

\begin{figure}[htb]
\begin{center}
\includegraphics[width=0.7\textwidth]{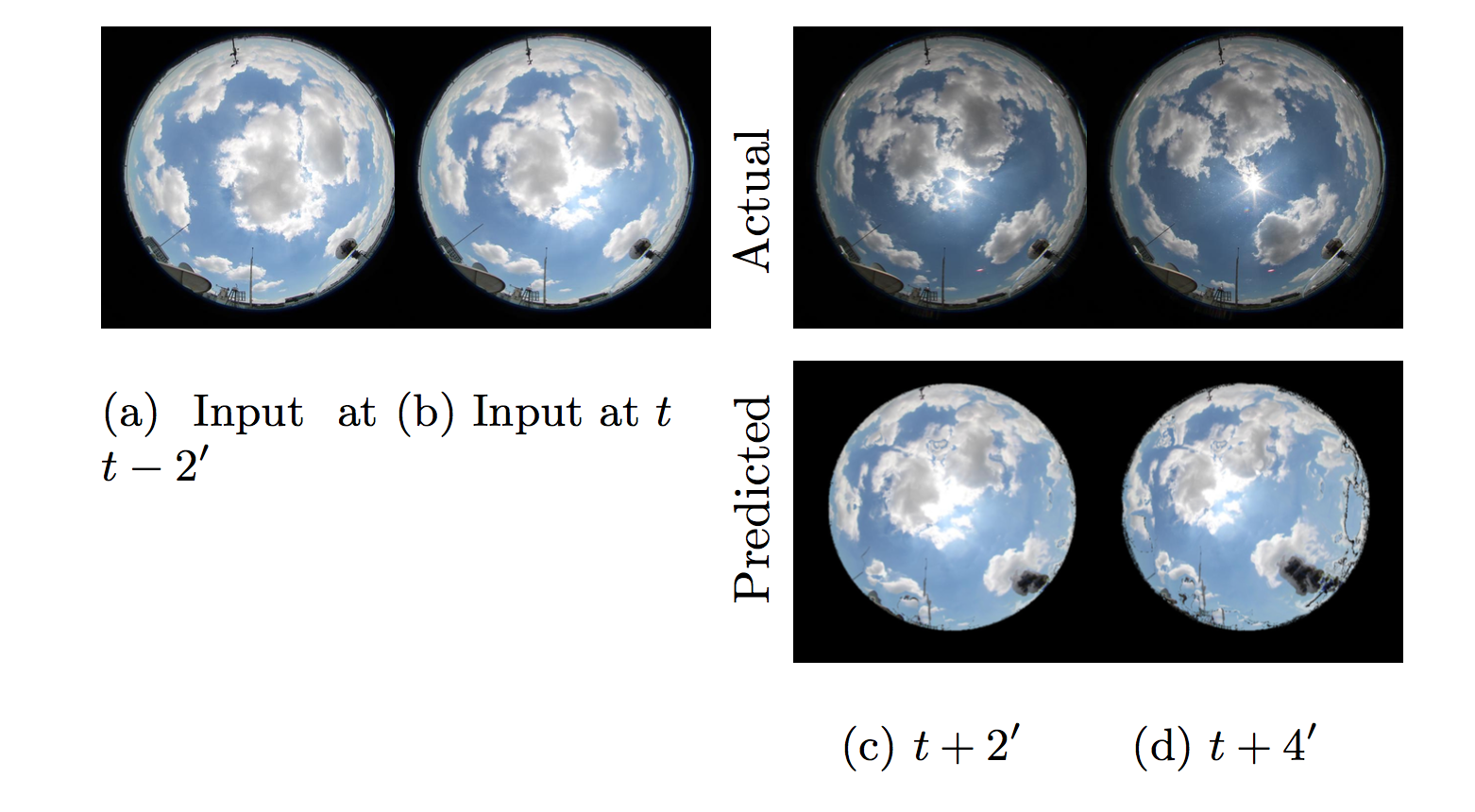}
\caption{Prediction of sky/cloud image using optical flow technique.
\label{fig:comb-example}}
\end{center}
\end{figure}

Using the images captured at $t$ and $t-2$ minutes, we estimate the horizontal and vertical translation. Under the assumption that the flow of cloud motion vectors for the successive $t+2$ minutes is similar to that of previous frames, we estimate the future $t+2$ minutes frame, and subsequently the $t+4$ minutes frame. Figure~\ref{fig:comb-example} illustrates this. We obtain a forecast accuracy of $70\%$ for a prediction lead time upto $6$ minutes.

In future work, we plan to use our proposed methodology of estimating solar irradiance on this predicted sky/cloud image. This will enable us to provide more stable and reliable forecasts of solar irradiance.

\subsection{Scope for Improvement}
%This paper proposes an empirical model to estimate the solar irradiance from ground-based cameras. However, t
There is still scope for improvement in our approach.  First, we use \emph{JPEG} images instead of uncompressed \emph{RAW} images for the computation of scene luminance. The \emph{JPEG} compression algorithm introduces non-linearities in the pixel values, which affects the process of estimating solar irradiance.  We can generate more consistent results by using only \emph{RAW} format images. Nevertheless, we still use \emph{JPEG} images, as they have a significantly smaller size, which is more practical from an operational point of view. In contrast, uncompressed \emph{RAW} images are much larger in size, which makes it impossible to capture and store \emph{RAW} images at short intervals due to the  processing requirements. In our future work, we intend to explore the use of \emph{RAW} format images for the computation of solar irradiance values from sky cameras. 

\begin{figure}[htb]
\begin{center}
\includegraphics[width=0.45\textwidth]{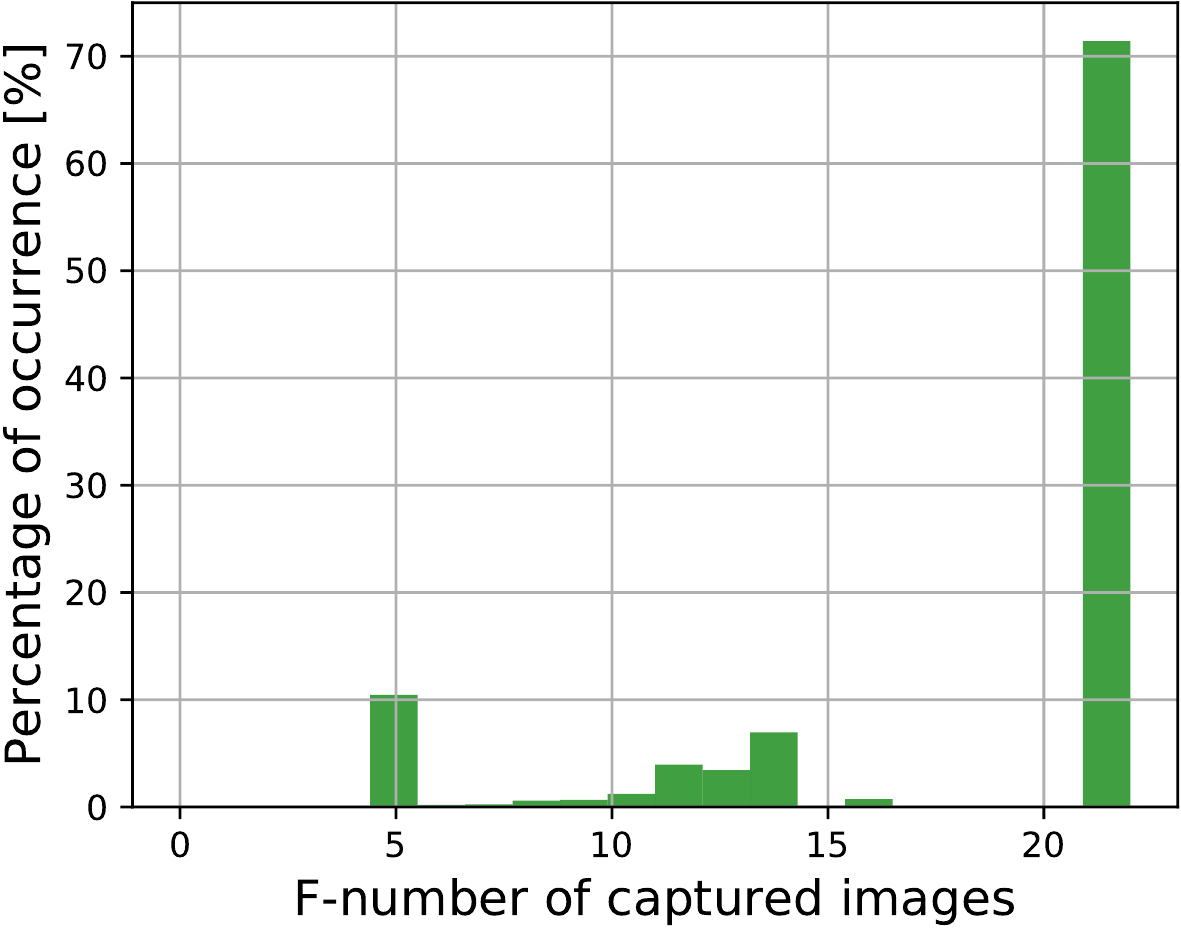}
\caption{Distribution of F-numbers of the WAHRSIS images used to derive the proposed model.
\label{fig:fn_dist}}
\end{center}
\end{figure}

Secondly, our captured images have a wide range of camera settings with varying shutter speed, ISO and aperture values. This is disadvantageous because the relationship between pixel value and camera aperture value becomes non-linear for larger F-numbers. The relationship deviates from linearity for F-numbers above $4.0$~\citep{hiscocks2011measuring}.  Figure~\ref{fig:fn_dist} depicts the wide range of F-numbers in the captured images used in deriving our proposed model. We observe that a significant percentage of images have large F-numbers, % greater than F $4.0$, 
where the non-linearity sets in. This can be solved by using the aperture priority mode of the sky camera, wherein the F-number is fixed, and the exposure time varies dynamically to match the lighting conditions of the scene.

\section{Conclusions \& Future work}
\label{sec:CLS}

\label{sec:conclusion}
We presented a method for estimating the rapid fluctuations of the solar irradiance using the luminance of images taken by a whole sky imager. We are able to estimate the rapid short-term variations, which significantly improves on the state-of-the-art. This approach is of interest for solar energy generation, because these variations cause a sudden decrease in the electricity output from solar panels. Short-term predictions of such ramp-downs are needed to maintain the stability of the power grid. 

Combining our solar irradiance estimation approach with cloud movement tracking in the input images could ultimately lead to better irradiance predictions. Such information on rapid fluctuations of solar irradiance can assist in establishing a high-reliability solar energy generation system. We also plan to explore methodologies from time-series modelling~\citep{dev2018solar} to predict solar irradiance.

\section{Code availability}
The source code of all simulations in this paper is available at \url{https://github.com/Soumyabrata/estimate-solar-irradiance}.

\appendix

\section{Derivation of Normalization Factor}
Let ${a_1, a_2, \ldots, a_t}$ be the weather station records for $t$ number of time stamps. The luminance values computed for each of the corresponding weather station points are represented by ${b_1, b_2, \ldots, b_t}$. We attempt to estimate the conversion factor $x$, such that the objective function $f(x)$ representing the inter-sample distances between weather station and computed luminance value is minimized.  We represent objective function $f(x)$ as:
\begin{align*}
   f(x) = \sum_{i=1}^{t} (xb_i - a_i)^2
\end{align*}

We compute the derivative $f'(x)$, and equate it to $0$, and the normalization factor $x$ is found as:
\begin{equation*}
    x = \frac{\sum_{i=1}^{t} a_{i}b_{i} }{\sum_{i=1}^{t} b_i^2}.
\end{equation*}

\begin{acknowledgements}
This research is funded by the Defence Science and Technology Agency (DSTA), Singapore. The  ADAPT  Centre  for  Digital  Content  Technology  is  funded  under  the  SFI Research Centres Programme (Grant 13/RC/2106) and is co-funded under the European Regional Development Fund.
\end{acknowledgements}

%% REFERENCES

%% The reference list is compiled as follows:

\bibliographystyle{copernicus}

\begin{thebibliography}{41}
\providecommand{\natexlab}[1]{#1}
\providecommand{\url}[1]{{\tt #1}}
\providecommand{\urlprefix}{URL }
\expandafter\ifx\csname urlstyle\endcsname\relax
  \providecommand{\doi}[1]{doi:\discretionary{}{}{}#1}\else
  \providecommand{\doi}{doi:\discretionary{}{}{}\begingroup
  \urlstyle{rm}\Url}\fi

\bibitem[{flo()}]{flow-pgm}
Optical Flow Matlab/C++ Code,
  \url{https://people.csail.mit.edu/celiu/OpticalFlow/}, accessed: 2019-08-09.

\bibitem[{nea()}]{nea-weather}
{NEA} | Weather, \url{http://www.nea.gov.sg/weather-climate/forecasts},
  accessed: 2019-08-09.

\bibitem[{Alonso-Montesinos and Batlles(2015)}]{alonso2015}
Alonso-Montesinos, J. and Batlles, F.~J.: The use of a sky camera for solar
  radiation estimation based on digital image processing, Energy, 90, Part 1,
  377--386, 2015.

\bibitem[{Alsadi and Nassar(2017)}]{alsadi2017estimation}
Alsadi, S.~Y. and Nassar, Y.~F.: Estimation of Solar Irradiance on Solar
  Fields: An Analytical Approach and Experimental Results, IEEE Transactions on
  Sustainable Energy, 8, 1601--1608, \doi{10.1109/TSTE.2017.2697913}, 2017.

\bibitem[{Baharin et~al.(2016)Baharin, Abdul~Rahman, Hassan, and
  Gan}]{baharin2016short}
Baharin, K.~A., Abdul~Rahman, H., Hassan, M.~Y., and Gan, C.~K.: Short-term
  forecasting of solar photovoltaic output power for tropical climate using
  ground-based measurement data, Journal of renewable and sustainable energy,
  8, 053\,701, 2016.

\bibitem[{Bristow and Campbell(1984)}]{BCmodel}
Bristow, K.~L. and Campbell, G.~S.: On the relationship between incoming solar
  radiation and daily maximum and minimum temperature, Agricultural and forest
  meteorology, 31, 159--166, 1984.

\bibitem[{Chu et~al.(2015)Chu, Urquhart, Gohari, Pedro, Kleissl, and
  Coimbra}]{chu2015short}
Chu, Y., Urquhart, B., Gohari, S.~M., Pedro, H.~T., Kleissl, J., and Coimbra,
  C.~F.: Short-term reforecasting of power output from a 48 {MWe} solar {PV}
  plant, Solar Energy, 112, 68--77, 2015.

\bibitem[{Dev et~al.(2014)Dev, Savoy, Lee, and Winkler}]{WAHRSIS}
Dev, S., Savoy, F.~M., Lee, Y.~H., and Winkler, S.: {WAHRSIS}: A low-cost,
  high-resolution whole sky imager with near-infrared capabilities, in: Proc.
  IS\&T/SPIE Infrared Imaging Systems, 2014.

\bibitem[{Dev et~al.(2015)Dev, Savoy, Lee, and Winkler}]{IGARSS2015a}
Dev, S., Savoy, F.~M., Lee, Y.~H., and Winkler, S.: Design of low-cost, compact
  and weather-proof whole sky imagers for {High-Dynamic-Range} captures, in:
  Proc. International Geoscience and Remote Sensing Symposium (IGARSS), pp.
  5359--5362, 2015.

\bibitem[{Dev et~al.(2017{\natexlab{e}})Dev, Lee, and Winkler}]{JSTARS2016}
Dev, S., Lee, Y.~H., and Winkler, S.: Color-Based Segmentation of Sky/Cloud
  Images From Ground-Based Cameras, IEEE Journal of Selected Topics in Applied
  Earth Observations and Remote Sensing, 10, 231--242, 2017{\natexlab{e}}.
  

\bibitem[{Dev et~al.(2016{\natexlab{b}})Dev, Savoy, Lee, and
  Winkler}]{IGARSS16_solar}
Dev, S., Savoy, F.~M., Lee, Y.~H., and Winkler, S.: Estimation of solar
  irradiance using ground-based whole sky imagers, in: Proc. International
  Geoscience and Remote Sensing Symposium (IGARSS), pp. 7236--7239,
  2016{\natexlab{b}}.

\bibitem[{Dev et~al.(2016{\natexlab{c}})Dev, Savoy, Lee, and
  Winkler}]{dev2016estimation}
Dev, S., Savoy, F.~M., Lee, Y.~H., and Winkler, S.: Estimation of solar
  irradiance using ground-based whole sky imagers, in: Proc. IEEE International
  Geoscience and Remote Sensing Symposium (IGARSS), pp. 7236--7239, IEEE,
  2016{\natexlab{c}}.

\bibitem[{Dev et~al.(2016{\natexlab{d}})Dev, Savoy, Lee, and Winkler}]{tencon}
Dev, S., Savoy, F.~M., Lee, Y.~H., and Winkler, S.: Short-term prediction of
  localized cloud motion using ground-based sky imagers, in: Proc. IEEE TENCON,
  2016{\natexlab{d}}.

\bibitem[{Dev et~al.(2016{\natexlab{e}})Dev, Wen, Lee, and Winkler}]{GRSM2016}
Dev, S., Wen, B., Lee, Y.~H., and Winkler, S.: Ground-Based Image Analysis: A
  Tutorial on Machine-Learning Techniques and Applications, IEEE Geoscience and
  Remote Sensing Magazine, 4, 79--93, 2016{\natexlab{e}}.

\bibitem[{Dev et~al.(2017)Dev, Manandhar, Lee, and Winkler}]{dev2017study}
Dev, S., Manandhar, S., Lee, Y.~H., and Winkler, S.: Study of clear sky models
  for Singapore, in: Progress in Electromagnetics Research Symposium-Fall
  (PIERS-FALL), 2017, pp. 1418--1420, IEEE, 2017.

\bibitem[{Dev et~al.(2018{\natexlab{a}})Dev, AlSkaif, Hossari, Godina, Louwen,
  and van Sark}]{dev2018solar}
Dev, S., AlSkaif, T., Hossari, M., Godina, R., Louwen, A., and van Sark, W.:
  Solar Irradiance Forecasting Using Triple Exponential Smoothing, in: 2018
  International Conference on Smart Energy Systems and Technologies (SEST), pp.
  1--6, IEEE, 2018{\natexlab{a}}.

\bibitem[{Dev et~al.(2018{\natexlab{b}})Dev, Savoy, Lee, and
  Winkler}]{dev2018high}
Dev, S., Savoy, F.~M., Lee, Y.~H., and Winkler, S.: High-dynamic-range imaging
  for cloud segmentation, Atmospheric Measurement Techniques, 11, 2041--2049,
  \doi{10.5194/amt-11-2041-2018},
  \urlprefix\url{https://www.atmos-meas-tech.net/11/2041/2018/},
  2018{\natexlab{b}}.

\bibitem[{Donatelli and Campbell(1998)}]{DCmodel}
Donatelli, M. and Campbell, G.~S.: A simple model to estimate global solar
  radiation, in: Proc. 5th European society of agronomy congress, pp. 133--134,
  1998.

\bibitem[{Feng et~al.(2018)Feng, Cui, Hodge, Lu, Hamann, and
  Zhang}]{feng2018unsupervised}
Feng, C., Cui, M., Hodge, B., Lu, S., Hamann, H., and Zhang, J.: Unsupervised
  Clustering-Based Short-Term Solar Forecasting, IEEE Transactions on
  Sustainable Energy, pp. 1--1, \doi{10.1109/TSTE.2018.2881531}, 2018.

\bibitem[{Hargreaves and Samani(1985)}]{HSmodel}
Hargreaves, G.~H. and Samani, Z.~A.: Reference crop evapotranspiration from
  temperature, Applied engineering in agriculture, 1, 96--99, 1985.

\bibitem[{Hiscocks and Eng(2011)}]{hiscocks2011measuring}
Hiscocks, P.~D. and Eng, P.: Measuring Luminance with a digital camera, Syscomp
  Electronic Design Limited, 2011.

\bibitem[{Hunt et~al.(1998)Hunt, Kuchar, and Swanton}]{Huntmodel}
Hunt, L.~A., Kuchar, L., and Swanton, C.~J.: Estimation of solar radiation for
  use in crop modelling, Agricultural and Forest Meteorology, 91, 293--300,
  1998.

\bibitem[{Huo and Lu(2012)}]{Huo2012}
Huo, J. and Lu, D.: Comparison of cloud cover from all-sky imager and
  meteorological observer, Journal of Atmospheric and Oceanic Technology, 29,
  1093--1101, 2012.

\bibitem[{Jang et~al.(2016)Jang, Bae, Park, and Sung}]{jang2016solar}
Jang, H.~S., Bae, K.~Y., Park, H., and Sung, D.~K.: Solar Power Prediction
  Based on Satellite Images and Support Vector Machine, IEEE Transactions on
  Sustainable Energy, 7, 1255--1263, \doi{10.1109/TSTE.2016.2535466}, 2016.

\bibitem[{Jiang et~al.(2017)Jiang, Long, Zhang, and Song}]{jiang2017day}
Jiang, Y., Long, H., Zhang, Z., and Song, Z.: Day-Ahead Prediction of Bihourly
  Solar Radiance With a Markov Switch Approach, IEEE Transactions on
  Sustainable Energy, 8, 1536--1547, \doi{10.1109/TSTE.2017.2694551}, 2017.

\bibitem[{Lautenbacher(2006)}]{lautenbacher2006global}
Lautenbacher, C.~C.: The global earth observation system of systems: Science
  serving society, Space Policy, 22, 8--11, 2006.

\bibitem[{Lef\`{e}vre et~al.(2014)Lef\`{e}vre, Blanc, Espinar, Gschwind,
  M\'{e}nard, Ranchin, Wald, Saboret, Thomas, and Wey}]{HelioClim2014}
Lef\`{e}vre, M., Blanc, P., Espinar, B., Gschwind, B., M\'{e}nard, L., Ranchin,
  T., Wald, L., Saboret, L., Thomas, C., and Wey, E.: The HelioClim-1 Database
  of Daily Solar Radiation at Earth Surface: An Example of the Benefits of
  GEOSS Data-CORE, IEEE Journal of Selected Topics in Applied Earth
  Observations and Remote Sensing, 7, 1745--1753, 2014.

\bibitem[{Lorenz et~al.(2009)Lorenz, Hurka, Heinemann, and Beyer}]{solar_PV}
Lorenz, E., Hurka, J., Heinemann, D., and Beyer, H.~G.: Irradiance Forecasting
  for the Power Prediction of Grid-Connected Photovoltaic Systems, IEEE Journal
  of Selected Topics in Applied Earth Observations and Remote Sensing, 2,
  2--10, 2009.

\bibitem[{Mueller et~al.(2004)Mueller, Dagestad, Ineichen,
  Schroedter-Homscheidt, Cros, Dumortier, Kuhlemann, Olseth, Piernavieja, Reise
  et~al.}]{mueller2004rethinking}
Mueller, R., Dagestad, K.-F., Ineichen, P., Schroedter-Homscheidt, M., Cros,
  S., Dumortier, D., Kuhlemann, R., Olseth, J., Piernavieja, G., Reise, C.,
  et~al.: Rethinking satellite-based solar irradiance modelling: The {SOLIS}
  clear-sky module, Remote sensing of Environment, 91, 160--174, 2004.

\bibitem[{Ouarda et~al.(2016)Ouarda, Charron, Marpu, and Chebana}]{SEVIRI2016}
Ouarda, T. B. M.~J., Charron, C., Marpu, P.~R., and Chebana, F.: The
  Generalized Additive Model for the Assessment of the Direct, Diffuse, and
  Global Solar Irradiances Using {SEVIRI} Images, With Application to the
  {UAE}, IEEE Journal of Selected Topics in Applied Earth Observations and
  Remote Sensing, 9, 1553--1566, 2016.

\bibitem[{Pagano and Durham(1993)}]{pagano1993moderate}
Pagano, T.~S. and Durham, R.~M.: Moderate resolution imaging spectroradiometer
  ({MODIS}), in: Proc. Sensor Systems for the Early Earth Observing System
  Platforms, vol. 1939, pp. 2--17, International Society for Optics and
  Photonics, 1993.

\bibitem[{Poynton(2003)}]{Poynton03}
Poynton, C.: Digital Video and {HDTV} Algorithms and Interfaces, Morgan
  Kaufmann Publishers Inc., 2003.

\bibitem[{Rizwan et~al.(2012)Rizwan, Jamil, and
  Kothari}]{rizwan2012generalized}
Rizwan, M., Jamil, M., and Kothari, D.~P.: Generalized Neural Network Approach
  for Global Solar Energy Estimation in India, IEEE Transactions on Sustainable
  Energy, 3, 576--584, \doi{10.1109/TSTE.2012.2193907}, 2012.

\bibitem[{Savoy et~al.(2016)Savoy, Dev, Lee, and Winkler}]{IGARSS16_calib}
Savoy, F.~M., Dev, S., Lee, Y.~H., and Winkler, S.: Geo-referencing and stereo
  calibration of ground-based whole sky imagers using the sun trajectory, in:
  Proc. International Geoscience and Remote Sensing Symposium (IGARSS), 2016.

\bibitem[{Scaramuzza et~al.(2006)Scaramuzza, Martinelli, and
  Siegwart}]{scaramuzza2006toolbox}
Scaramuzza, D., Martinelli, A., and Siegwart, R.: A toolbox for easily
  calibrating omnidirectional cameras, in: Proc. IEEE/RSJ International
  Conference on Intelligent Robots and Systems, pp. 5695--5701, IEEE, 2006.

\bibitem[{Shakya et~al.(2017)Shakya, Michael, Saunders, Armstrong, Pandey,
  Chalise, and Tonkoski}]{shakya2017Solar}
Shakya, A., Michael, S., Saunders, C., Armstrong, D., Pandey, P., Chalise, S.,
  and Tonkoski, R.: Solar Irradiance Forecasting in Remote Microgrids Using
  Markov Switching Model, IEEE Transactions on Sustainable Energy, 8, 895--905,
  \doi{10.1109/TSTE.2016.2629974}, 2017.

\bibitem[{Silva and Souza-Echer(2016)}]{Silva2016}
Silva, A.~A. and Souza-Echer, M.~P.: Ground-based observations of clouds
  through both an automatic imager and human observation, Meteorological
  Applications, 23, 150--157, 2016.

\bibitem[{SMPTE(1993)}]{smpte1993rp}
SMPTE, R.: {RP} 177-1993, Derivation of Basic Television Color Equations, 1993.

\bibitem[{Yang and Chen(2015)}]{yang2015expanding}
Yang, D. and Chen, N.: Expanding Existing Solar Irradiance Monitoring Network
  Using Entropy, IEEE Transactions on Sustainable Energy, 6, 1208--1215,
  \doi{10.1109/TSTE.2015.2421734}, 2015.

\bibitem[{Yang et~al.(2012)Yang, Jirutitijaroen, and
  Walsh}]{dazhi2012estimation}
Yang, D., Jirutitijaroen, P., and Walsh, W.~M.: The estimation of clear sky
  global horizontal irradiance at the equator, Energy Procedia, 25, 141--148,
  2012.

\bibitem[{Yuan et~al.(2016)Yuan, Lee, Meng, and Ong}]{Yuan_TGRS15}
Yuan, F., Lee, Y.~H., Meng, Y.~S., and Ong, J.~T.: Water Vapor Pressure Model
  for Cloud Vertical Structure Detection in Tropical Region, IEEE Transactions
  on Geoscience and Remote Sensing, 54, 5875--5883, 2016.

\end{thebibliography}

\end{document}